\documentclass[aps,pra,twocolumn,showpacs]{revtex4}

\usepackage{hhline}
\usepackage{graphicx}
\usepackage{bbm}
\usepackage{amsfonts}
\usepackage{theorem}
\usepackage{amssymb}
\usepackage{amsmath}

\newcommand{\diag}{\mathop{\mbox{diag}}\nolimits}
\newcommand{\rank}{\mathop{\mbox{rank}}\nolimits}

\begin{document}
\title{\begin{flushright}\begin{small}UWthPh--2005--23\end{small}\end{flushright}
  Maximizing nearest neighbour entanglement in
  finitely correlated qubit--chains}

\author{Beatrix C. Hiesmayr$^1$, M\'aty\'as Koniorczyk$^{2,3}$ and Heide
  Narnhofer$^1$}
\address{$^1$Institut f\"ur Theoretische Physik, Universit\"at Wien, Boltzmanngasse 5, 1090 Vienna, Austria}
\address{$^2$Research Institute for Solid State Physics and Optics, Hungarian Academy of Sciences, 1525 Budapest, P.O. Box 49, Hungary}
\address{$^3$Institute of Physics, University of P\'ecs, 7624 P\'ecs,
  Ifj\'us\'ag \'utja 6, Hungary}
\date{\today}
\begin{abstract}
We consider translationally invariant states of an infinite one
dimensional chain of qubits or spin-$\frac{1}{2}$ particles. We
maximize the entanglement shared by nearest neighbours via a
variational approach based on finitely correlated states. We find an
upper bound of nearest neighbour concurrence equal to ${\cal
C}=0.434095$ which is $0.09\%$ away from the bound ${\cal
C}_\text{W}=0.434467$ obtained by a completely different procedure.
The obtained state maximizing nearest neighbour entanglement seems
to approximate the maximally entangled mixed states (MEMS). Further
we investigate in detail several other properties of the so obtained
optimal state. 
\end{abstract}
 \maketitle

%\tableofcontents

\section{Introduction}

The understanding of entanglement in a multipartite quantum system is
a central problem of contemporary quantum mechanics, also spreading to
statistical and solid state physics. A multipartite quantum system
cannot exhibit arbitrary entanglement properties, and the restrictions
are far from being straightforward. This fact has several implications
on the properties of spin chains and spin lattices, the typical
subjects of statistical and solid state physics. This became apparent
along with the recent developments of density matrix renormalization
group (DMRG) techniques (consult ~\cite{DMRG} for a recent review).

In particular, if one considers a system of two subsystems in a
maximally entangled state, neither of the subsystems can, of course,
be entangled with anything else. In a system of many quantum bits
(qubits, or spin-$\frac{1}{2}$ particles) the limitations on the
entanglement of the qubit-pairs are quantified by the
Coffman-Kundu-Wootters (CKW) inequalities~\cite{CKW}, for more than
three qubits this long standing conjecture was recently proven
\cite{CKWproof}. Finding a quantum state with prescribed pairwise
entanglement between each pair is therefore not always possible, and
it is a rather involved task~\cite{Buzek}.

In this paper we consider an infinite one dimensional chain of qubits where
each qubit is entangled at least with its two nearest neighbours. We impose the
constraint of translation invariance: the state should be invariant under all
transformations that shift each qubit from its original position $i$ to $i+n$
for some integer $n$. Our main goal is to find the maximal possible achievable
entanglement of the nearest neighbours, and study the properties of the so
arising chain.  This problem is interesting mainly for two reasons. First,
because if the bound on nearest neighbour entanglement is optimal, then it
serves as a reference point for interpreting entanglement values obtained for
real physical systems (such as the antiferromagnetic Heisenberg
chain~\cite{ConnorWooters} or lines of ions in a trap which could be used for
quantum computations~\cite{CiracZoller}).  Second, because the results
contribute to the knowledge on the possible structures of distributed
entanglement in systems of (infinitely) many subsystems.

This question was also addressed by
Wootters~\cite{WoottersEntangledChains}.  Via a certain procedure he
succeeded in constructing such translation invariant entangled chains
in which the maximal achievable concurrence is ${\cal
  C}_{\text{W}}=0.434467$ which corresponds to a value of entanglement
of formation of $E_f=0.284934$ ebits. This value is below ${\cal
  C}_{\text{CKW}}=1/\sqrt{2}$, the limit that the CKW inequalities
would allow for, in the case when each quantum bit is maximally entangled
with the rest of the system, while the bipartite entanglement is
restricted to the nearest neighbours. The concurrence ${\cal C}_W$ is
conjectured to be an absolute bound, but this fact is not proven. A
challenging question in this context is if one can go beyond ${\cal
  C}_W$, or even reach ${\cal C}_{\text{CKW}}$.

In this paper we attack the problem in a different way, related to the DMRG
method. This latter is found to be a variational method in terms of the
so-called matrix product states (MPS)~\cite{FNW}, equivalent to a sequence of
entanglement swappings~\cite{DMRGinf}. The MPSs constitute a representation of
the pure state of a finite number of qubits. For each qubit of the system an
auxiliary finite dimensional Hilbert space is considered. The state is
described by projectors acting between the auxiliary space and the space of the
system under consideration. They are very suitable for approximating ground
states of Hamiltonians in a numerically efficient way where the approximation
lies on the dimensionality of the auxiliary system. The infinite chain can be
studied as a limit, assuming periodic boundary conditions. This attitude is
equivalent to a different formulation, termed finitely correlated states (FCS).
A pure state of the whole translationally invariant chain is encoded into a
state of an auxiliary system and a completely positive (CP) map. In this
framework, the density operator for any finite subset of the system can be
constructed by the successive application of the CP map, while the auxiliary
system models the rest of the system from the point of view of quantum
correlations.  All translational invariant states can be approximated in that
way with an accuracy which may depend on the dimension of the auxiliary system.

The entanglement distribution of such chains has been already studied in
Refs.~\cite{BHN1,BHN2} under the restriction of $2$ dimensional auxiliary
systems~\cite{endnote1}.

We shall use a specific subset of such FCS as an ansatz to maximize nearest
neighbour entanglement in a translationally invariant infinite chain of qubits.
This approach has several advantages. First, it provides us explicitly with a
well-defined pure state on the whole chain.
%Secondly, we can study another class than
%Ref.~\cite{WoottersEntangledChains} of nearest neighbour states of
%such an entangled chain.
Second, we can study a class of quantum states different to the one studied in
Ref.~\cite{WoottersEntangledChains}. Third, we can also investigate higher
correlations as next nearest neighbour entanglement and entanglement of one
qubit with the rest of the chain.

Thus we can study the entanglement distributed along the chain,
however, we focus mainly on maximizing nearest neighbour
entanglement and the properties of such an optimal state regarding
the entanglement of next nearest, and further neighbours. For two
dimensional auxiliary systems we have analytical results describing
the properties of entanglement of such an entangled chain and for
higher dimensions we have performed a numerical optimization. We
have found that the achievable nearest neighbours entanglement seems
to converge fast to Wootters' bound ${\cal C}_\text{W}$.

This paper is organized as follows. Section~\ref{sec:states} is devoted to the
construction of the states of translationally invariant finitely correlated
chains utilized in this paper. In Section~\ref{max} analytical results are
presented for the maximization of the nearest neighbour entanglement for the
case of low dimensional auxiliary Hilbert spaces, to give an insight into the
nature of the problem. In Section~\ref{numericaloptimization} our numerical
results are presented for higher dimensionalities, while in
Section~\ref{sec:properties} the properties of the so found optimal states are
discussed. In Section~\ref{sect:summary} our results are summarized, and
conclusions are drawn.

\section{Construction of the translation invariant entangled chain}
\label{sec:states}

We consider an infinite ensemble of qubits arranged along a line.
The first question we address is what we mean by the word ``state''
as applied to infinitely many qubits. We adopt the standard approach
described in e.g.~Ref.~\cite{bookBrotteliRobinson}. A state $\omega$
of the infinite chain is a functional that assigns to every finite
set of local operators $A_{[1,n]}=A_1\otimes A_2\dots\otimes A_n$ a
normalized density matrix $\omega(A_{[1,n]})$ describing the
properties of the of $n$ qubits. Moreover one demands that if one
considers a subset of local operators $A_{[1,k]}$ of the set
$A_{[1,n]}$ then the state $\omega(A_{[1,k]})$ has to be obtained by
taking the partial trace of $\omega(A_{[1,n]})$ over the qubits not
in $A_{[1,n]}$.

Before describing the rather mathematical construction in detail, we
outline first the idea behind it.
%The details of the construction are given in the next section, however, since
%it is rather mathematical we outline first the idea.
%Such a state
%$\omega(A_{[1,n]})$ of a string of $n$ qubits is then defined as an expectation
%over local operators $A_{[1,n]}=A_1\otimes A_2\dots\otimes A_n$ acting on a
%density matrix $\rho_{[1,n]}$ describing the properties of these $n$ qubits in
%such an infinite chain. Clearly, $A$ and $\rho_{[1]}={\rm
%Tr}_{[2,n]}\left(\rho_{[1,n]}\right)$ are elements of a $2\times 2$ dimensional
%Hilbert space which we can generally represent by $2\times 2$ complex matrices.
%
%Let us first investigate the density matrix of one qubit of the
%chain.
The trick is to describe the part of the chain in which one is not
interested in by a density matrix on an auxiliary Hilbert space ${\cal
  H}_B$, a local operator of the bounded operators $\cal{B}({\cal H}_B)$ and a completely
positive map which maps tensor products of such an auxiliary local
operator and the local operator of one qubit $A_i$ always back into
$\cal{B}({\cal H}_B)$. In this way the completely positive map ensures
that one only archives permitted states of a qubit or more qubits in
such an infinite chain. It is clear that the set of the permitted and thus
realized states of a qubit or more qubits of the chain can increase if
the dimension of the auxiliary system describing the ``rest of the
chain'' is increased. Further one has to carefully choose the
appropriate completely positive maps in order to study the class of
permitted one or more qubit states one is interested in.

Explicit examples for choices of the completely positive map for increasing
dimensions of the auxiliary system which maximize nearest neighbour
entanglement (our main goal) are then given in Sec.~\ref{max}, but let us first
proceed with the construction of the finitely correlated states in more detail.

\subsection{Construction of finitely correlated states
(FCS)}\label{fcs}

In the following we summarize the exact mathematical construction of
translationally invariant finitely correlated states according to
Ref.~\cite{FNW}. We denote by $\mathcal{A}_\mathbb{Z}$ an infinite
spin-chain, the spins at sites $i\in \mathbb{Z}$ being described by
the algebra $({\cal A})_i={\bf M}_2$ of $2\times 2$ complex matrices,
i.e. we describe spin-$\frac{1}{2}$ particles or generally qubits. The
infinite algebra $\mathcal{A}_\mathbb{Z}$ arises as a suitable limit
of the {\it local} tensor-product algebras ${\cal
  A}_{[-n,n]}:=\otimes_{j=-n}^n(\mathcal{A})_j$. Any state $\omega$
over $\mathcal{A}_\mathbb{Z}$ is specified by density matrices
$\rho_{[1,n]}$ defining the action of $\omega$ as an expectation over
local operators $A_{[1,n]}\in\mathcal{A}_{[1,n]}$:
\begin{eqnarray}
\label{fcs0} \omega(A_{[1,n]})&=& {\rm
Tr}_{[1,n]}\Bigl(\rho_{[1,n]}\,
A_1\otimes A_2\otimes \dots A_n \Bigr)\nonumber\\
&=&{\rm Tr}_{[1,n]}\Bigl(\rho_{[1,n]}\,A_{[1,n]}\Bigr) \ .
\end{eqnarray}
The $\rho_{[1,n]}$'s must satisfy the compatibility conditions, i.e.
acting locally on the $n+1$ qubit with a unity operator should give
the same expectation value
\begin{eqnarray}
\omega(A_{[1,n]})&=&\label{comp}
\hbox{Tr}_{n+1}\Bigl(\rho_{[1,n+1]}\,A_{[1,n]}\otimes
\mathbbm{1}_{n+1}\Bigl)\nonumber\\
&=&\hbox{Tr}_{[1,n]}\Bigl(\rho_{[1,n]}\,A_{[1,n]}\Bigr)\ .
\end{eqnarray}
Whereas, translation-invariance requires that doing no operation on
the first qubit should also do no change to the expectation value,
i.e. shifting the line of qubits by one (generally by an integer)
\begin{eqnarray}
\label{trans-inv}
\omega(A_{[1,n]})&=&\hbox{Tr}_{n+1}\Bigl(\rho_{[1,n+1]}\,
\mathbbm{1}_1\otimes A_{[2,n+1]}\Bigl)\nonumber\\
&=&\hbox{Tr}_{[1,n]}\Bigl(\rho_{[1,n]}\,A_{[1,n]}\Bigr)\ .
\end{eqnarray}
The class of translation-invariant finitely correlated states (FCS)
over $\mathcal{A}_{\mathbb{Z}}$ is defined by a triple
$(\mathcal{B},\rho,\mathbb{E})$ where $\mathcal{B}$ is a $b\times b$
matrix algebra $\mathcal{B}$,  $\rho_B\in\mathcal{B}$ a density
matrix and
$\mathbb{E}:\mathcal{A}\otimes\mathcal{B}\mapsto\mathcal{B}$ a
completely positive unital map, which in Kraus form reads
\begin{equation}
\label{Vop0} \mathbb{E}(A\otimes B)=\sum_j V_j(A\otimes B)
V_j^\dagger\ ,\quad
V_j:\mathbb{C}^2\otimes\mathbb{C}^b\mapsto\mathbb{C}^b\ ,
\end{equation}
with $A\in\mathcal{A}$ and $B\in\mathcal{B}$. Unitality means that
identities are preserved
\begin{equation}
\mathbb{E}(\mathbbm{1}_\mathcal{A}\otimes\mathbbm{1}_\mathcal{B})=\mathbbm{1}_\mathcal{B}\quad\textit{
unitality}\,.
\end{equation}
Let
$\mathbb{E}^{(1)}(A):=\mathbb{E}(A\otimes\mathbbm{1}_\mathcal{B})$;
this defines a completely positive map from $\mathcal{A}$ into
$\mathcal{B}$. Analogously, the recursive compositions
$\mathbb{E}^{(n)}:=\mathbb{E}\circ\Bigl({\rm id}_\mathcal{A}\otimes
\mathbb{E}^{(n-1)}\Bigr)$ are completely positive maps from
$\mathcal{A}_{[1,n]}$ into $\mathcal{B}$. Setting
\begin{widetext}
\begin{eqnarray}\label{st1}
\omega(A_{[1,n]})=\hbox{Tr}
\Bigl(\rho_{[1,n]}\,A_{[1,n]}\Bigr)&:=&{\rm Tr}_{{\cal B}}
\big(\rho_{B}\, \underbrace{\mathbb{E}\big[
A_1\otimes\underbrace{\mathbb{E}\big[A_2\otimes
\underbrace{\mathbb{E}\big[\dots\otimes\big[\underbrace{A_n\otimes\mathbbm{1}_{{\cal
B}}}_{\in{\cal B}}\big]\dots\big]}_{\in{\cal B}}\big]}_{\in{\cal
B}}\big]}_{\in{\cal B}}\big)
\nonumber\\
&=& {\rm Tr}_{{\cal{B}}}
\Bigl(\rho_B\,\mathbb{E}^{(n)}\Bigl(A_{[1,n]}\Bigr)\Bigr)\ ,
\end{eqnarray}
\end{widetext}
the r.h.s. recursively defines local density matrices $\rho_{[1,n]}$
over $\mathcal{A}_{[1,n]}$ and a total state $\omega$ on
$\mathcal{A}_\mathbb{Z}$. Translation invariance
condition  Eq.~(\ref{trans-inv}) can be formulated as
\begin{equation}
{\rm
Tr}_\mathcal{B}\left(\rho_B\,\mathbb{E}\left(\mathbbm{1}_{\mathcal{A}}\otimes
B\right)\right)={\rm Tr}_\mathcal{B}(\rho_B\,B)\quad\forall\,
B\in\mathcal{B}\,.
\end{equation}

\subsection{Constraints of unitality and translational invariance}\label{translationalinvariance}

Concretely, we choose $\mathcal{B}={\bf M}_b$ ($b\times b$ complex
matrices) and $\mathbb{E}$ like in~(\ref{Vop0}) but with just one
Kraus operator $V:\mathbb{C}^2\otimes\mathbb{C}^b\mapsto\mathbb{C}^b$.
(Note that this restriction does not decrease the generality, as the
introduction of further Kraus operators can be avoided by increasing
the dimensionality $b$.) This is such that $V\vert
a_i\otimes\psi\rangle=v_i\vert\psi\rangle$, $V^\dagger\vert\psi\rangle
=\sum_{i=1}^2\vert a_i\rangle\otimes v_i^\dagger\vert\psi\rangle$ and
with the one qubit operator $A=\sum a_{ij}\ \vert a_i\rangle\langle
a_j\vert$
\begin{eqnarray}
\label{1kraus} \mathbb{E}\Bigl(\vert a_i\rangle\langle
a_j\vert\otimes B\Bigr) = v_i\, B\, v_j^\dagger\ ,\quad
B\in\mathcal{B}\ ,
\end{eqnarray}
where $\vert a_{1,2}\rangle\in\mathbb{C}^2$ are orthonormal and
$v_{1,2}$ $b\times b$ matrices. In this notation the unitality and
translation invariance reads for the two matrices
\begin{eqnarray}\label{condition1}
&&v_1 v_1^\dagger+v_2 v_2^\dagger=\mathbbm{1}_\mathcal{B}\quad
unitality\\
\label{condition2} &&\sum_{i=1}^2v_i^\dagger\ \rho_B\
v_i=\rho_B\quad \text{\textit{translation invariance}}\,.
\end{eqnarray}
If there exists a unique $\rho$ fulfilling the previous condition, the
resulting translation--invariant \textsf{FCS} are pure states over
$\mathcal{A}_\mathbb{Z}$~\cite{FNW}, namely they cannot be decomposed as
mixtures of other states. These pure states can be interpreted as ground states
for appropriately constructed Hamiltonians of finite range~\cite{FNW}. It is
also shown that this class of states is dense in the set of all translation
invariant states~\cite{FNW2}.

The two conditions Eq.~(\ref{condition1}), Eq.~(\ref{condition2}) can as well
be interpreted in the context of open quantum systems. I.e. $\rho_B$ is the
state of some open quantum system where the $v$'s are the operation elements
for the quantum operation. These operation elements satisfy the well known
completeness relation (first condition) which leads to trace conservation of
the completely positive map, i.e. no information of the whole system ($B$ plus
environment) is lost. The second equation (condition 2) can then be interpreted
as searching for states which are invariant under these interactions.

\subsection{Density matrices for a subset of qubits in the chain}

Let us first consider the state of one qubit in the chain, defined
in Eq.(\ref{st1})
\begin{eqnarray}
\omega(A)&=&{\rm Tr}_A\big(\rho_{[1]}\ A\otimes
\mathbbm{1}_B\big):={\rm
Tr}_B\big(\rho_B\ \mathbb{E}(A\otimes \mathbbm{1}_B)\big)\nonumber\\
&=&{\rm Tr}_B\left(\rho_B\ \mathbb{E} \left(\sum a_{ij}\ \vert
a_i\rangle\langle a_j\vert\otimes \mathbbm{1}_B
\right)\right)\nonumber\\
&=& {\rm
Tr}_B\big(\rho_B \sum a_{ij}\ v_i \mathbbm{1}_B v_j^\dagger\big)\nonumber\\
&=&{\rm Tr}_{B}\left(\sum a_{ij}\ v_j^\dagger \rho_B v_i\
\mathbbm{1}_B\right)\nonumber\\
&=& {\rm Tr}_{A\otimes B}\left(\sum a_{ij}\ {\rm
Tr}_B\left(v_j^\dagger \rho_B v_i\right)\ \vert a_i\rangle\langle
a_j\vert\otimes \mathbbm{1}_B\right)\nonumber\\
\end{eqnarray}
where we have used the cyclic property of the trace -- ${\rm
  Tr}(XY)={\rm Tr}(YX)$ -- and the fact that the trace operation is
invariant under the map $\mathbb{F}:{\cal B}\rightarrow{\cal A}\otimes
{\cal B}$ dual to $\mathbb{E}:{\cal A}\otimes {\cal B}\rightarrow
{\cal B}$.  This is defined by
\begin{eqnarray}
\mathbb{F}(B):=\sum_{i,j=2}^2 \vert a_j\rangle\langle a_i\vert
\otimes v_i^\dagger B v_j\;. \end{eqnarray} Hence comparing the last
term with the first one we find that the density matrix of a qubit
in the chain is given by
\begin{eqnarray}
\rho_{[1]}&=&\rho_1=\sum_{i,j}^2 \vert a_j\rangle\langle a_i\vert\
{\rm Tr}\big(v_j^\dagger\, \rho_B\,
v_i\big)\nonumber\\
&=&\left(\begin{array}{cc} {\rm Tr}\big(v_1^\dagger\, \rho_B\,
v_1\big)&{\rm Tr}\big(v_1^\dagger\, \rho_B\, v_2\big)\\ {\rm
Tr}\big(v_2^\dagger\, \rho_B\, v_1\big)&{\rm Tr}\big(v_2^\dagger\,
\rho_B\, v_2\big)\end{array}\right)\;.
\end{eqnarray}
Now let us investigate the nearest neighbour state, i.e.
Eq.(\ref{st1}) by setting $n=2$, thus we have $A_{[1,2]}=A_1\otimes
A_2$ and
\begin{equation}
\hbox{Tr} \Bigl(\rho_{[1,2]}\,A_1\otimes A_2\Bigr):={\rm Tr}_{B}
\Bigl(\rho_B\,\mathbb{E}\Bigl(A_1\otimes\mathbb{E}\Bigl(A_2\otimes\mathbbm{1}_B
\Bigr)\Bigr)\ .
\end{equation}
Again using the properties of the trace-operation, the action of
$\mathbb{E}$ becomes the action of its dual map $\mathbb{F}$
\begin{equation}
\label{expl} \hbox{Tr} \Bigl(\rho_{[1,2]}\,A_1\otimes
A_2\Bigr):={\rm Tr}_{A\otimes B}
\Bigl(\mathbb{F}(\rho_B)\,A_1\otimes
\mathbb{E}\Bigl(A_2\otimes\mathbbm{1}_B \Bigr)\Bigr)\ .
\end{equation}

 This provides a state $\rho_{\mathcal{A}\otimes\mathcal{B}}:=
\mathbb{F}(\rho_B)=V^\dagger\,\rho_B\,V$ on
$\mathcal{A}\otimes\mathcal{B}$:
\begin{equation}
\label{qbc10} \rho_{\mathcal{A}\otimes\mathcal{B}}=
\sum_{s,t=1}^2\vert s\rangle\langle t\vert\,\otimes\,
v^\dagger_s\,\rho_B\, v_t =\begin{pmatrix} v_1^\dagger\,\rho_B\,
v_1&v_1^\dagger\,\rho_B\, v_2\cr v_2^\dagger\,\rho_B\,
v_1&v_2^\dagger\,\rho_B\, v_2
\end{pmatrix}
\end{equation}
which encodes the properties of all the correlations between one site
with the rest of the whole chain.

The right hand side of~Eq.(\ref{expl}) reads
$\hbox{Tr}_{\mathcal{A}\otimes\mathcal{B}}
\Bigl(\rho_{\mathcal{A}\otimes\mathcal{B}}\,A_1\otimes\mathbb{E}
\left(A_2\otimes \mathbbm{1}_\mathcal{B}\right)\Bigr)$, by turning
${\rm id}_\mathcal{A}\otimes\mathbb{E}$ into its dual,
nearest-neighbours states arise as $\rho_{12}:=\rho_{[1,2]}
=\hbox{Tr}_\mathcal{B}\Bigl({\rm id}_\mathcal{A}\otimes\mathbb{F}
(\rho_{\mathcal{A}\otimes\mathcal{B}})\Bigr)$ which reads
\begin{eqnarray}\label{rho12R}
\rho_{12}&=&\sum_{ij=1}^2 \vert i\rangle\langle j\vert\otimes
\begin{pmatrix}
R_{1ij1}&R_{1ij2}\cr
R_{2ij1}&R_{2ij2}\end{pmatrix}%\\
%\label{qbc11} &=&
%\begin{pmatrix}
%{\rm Tr}(v_1^\dagger v_1^\dagger\;\rho\;v_1 v_1)&{\rm
%Tr}(v_1^\dagger v_1^\dagger\;\rho\;v_1 v_2)&{\rm Tr}(v_1^\dagger
%v_1^\dagger\;\rho\;v_2 v_1)&{\rm Tr}(v_1^\dagger
%v_1^\dagger\;\rho\;v_2 v_2)\cr {\rm Tr}(v_2^\dagger
%v_1^\dagger\;\rho\;v_1 v_1)&{\rm Tr}(v_2^\dagger
%v_1^\dagger\;\rho\;v_1 v_2)&{\rm Tr}(v_2^\dagger
%v_1^\dagger\;\rho\;v_2 v_1)&{\rm Tr}(v_2^\dagger
%v_1^\dagger\;\rho\;v_2 v_2)\cr {\rm Tr}(v_1^\dagger
%v_2^\dagger\;\rho\;v_1 v_1)&{\rm Tr}(v_1^\dagger
%v_2^\dagger\;\rho\;v_1 v_2)&{\rm Tr}(v_1^\dagger
%v_2^\dagger\;\rho\;v_2 v_1)&{\rm Tr}(v_1^\dagger
%v_2^\dagger\;\rho\;v_2 v_2)\cr {\rm Tr}(v_2^\dagger
%v_2^\dagger\;\rho\;v_1 v_1)&{\rm Tr}(v_2^\dagger
%v_2^\dagger\;\rho\;v_1 v_2)&{\rm Tr}(v_2^\dagger
%v_2^\dagger\;\rho\;v_2 v_1)&{\rm Tr}(v_2^\dagger
%v_2^\dagger\;\rho\;v_2 v_2)\cr
%\end{pmatrix}
\ ,
\end{eqnarray}
where $R_{ijlm}={\rm Tr}(v_i^\dagger v_j^\dagger\; \rho_B\; v_l
v_m)$.

In general local density matrices are constructed by
\begin{equation}
\label{qbc5} \rho_{[1,n]}\;=\;\sum_{{\bf s},{\bf t}}\vert{{\bf
s}}\rangle \langle{\bf t}\vert\; {\rm Tr}(v_{{\bf
s}}^\dagger\,\rho_B\,v_{{\bf t}})\ ,
\end{equation}
where $\vert{{\bf s}}\rangle =\vert s_1\otimes s_2\otimes\cdots s_n
\rangle$, $v_{\bf t}:=v_{t_1}\cdots v_{t_n}$.

\section{Maximizing nearest neighbour entanglement}
\label{max}

\subsection{General discussion for an optimal choice of
$v_1$}\label{generaldiscussionv1}

Our main goal is to maximize the nearest neighbour entanglement of
state Eq.~(\ref{rho12R})
\begin{widetext}
\begin{eqnarray}\label{gnnentanglement} \rho_{12}&=&
\begin{pmatrix}
{\rm Tr}(v_1^\dagger v_1^\dagger\;\rho_B\;v_1 v_1)&{\rm
Tr}(v_1^\dagger v_1^\dagger\;\rho_B\;v_1 v_2)&{\rm Tr}(v_1^\dagger
v_1^\dagger\;\rho_B\;v_2 v_1)&{\rm Tr}(v_1^\dagger
v_1^\dagger\;\rho_B\;v_2 v_2)\cr {\rm Tr}(v_2^\dagger
v_1^\dagger\;\rho_B\;v_1 v_1)&{\rm Tr}(v_2^\dagger
v_1^\dagger\;\rho_B\;v_1 v_2)&{\rm Tr}(v_2^\dagger
v_1^\dagger\;\rho_B\;v_2 v_1)&{\rm Tr}(v_2^\dagger
v_1^\dagger\;\rho_B\;v_2 v_2)\cr {\rm Tr}(v_1^\dagger
v_2^\dagger\;\rho_B\;v_1 v_1)&{\rm Tr}(v_1^\dagger
v_2^\dagger\;\rho_B\;v_1 v_2)&{\rm Tr}(v_1^\dagger
v_2^\dagger\;\rho_B\;v_2 v_1)&{\rm Tr}(v_1^\dagger
v_2^\dagger\;\rho_B\;v_2 v_2)\cr {\rm Tr}(v_2^\dagger
v_2^\dagger\;\rho_B\;v_1 v_1)&{\rm Tr}(v_2^\dagger
v_2^\dagger\;\rho_B\;v_1 v_2)&{\rm Tr}(v_2^\dagger
v_2^\dagger\;\rho_B\;v_2 v_1)&{\rm Tr}(v_2^\dagger
v_2^\dagger\;\rho_B\;v_2 v_2)\cr
\end{pmatrix}\ .
\end{eqnarray}
\end{widetext} To do so we have to choose appropriate matrices $v_1,
v_2$ satisfying the condition in Eq.~(\ref{condition1}) and then to
derive a unique $\rho_B$ satisfying the condition in
Eq.~(\ref{condition2}). This then defines the nearest neighbour
entanglement given by the density matrix above.

\begin{center}
What are appropriate choices for $v_1$?
\end{center}

Obviously the entanglement shared by two neighbouring qubits in a
chain cannot be maximal, e.g. one of the four pure entangled Bell type
states ($|\phi^{\pm}\rangle=\frac{1}{\sqrt{2}}\lbrace
|00\rangle\pm|11\rangle\rbrace,
|\psi^{\pm}\rangle=\frac{1}{\sqrt{2}}\lbrace
|01\rangle\pm|01\rangle\rbrace$). In this case every other qubit has
to be disentangled with that pair. Therefore the reduced state of the
infinite chain we are looking for is not pure. On the other hand it
should be far away from the tracial state (i.e. the complete mixture)
as well. It seems to be plausible though not conclusive that the
reduced state vanishes on some subspace. Considering the basis states
$|00\rangle, |11\rangle, |01\rangle, |10\rangle$ we notice that if we
choose $|01\rangle$ states then because of translation invariance we
need as well $|10\rangle$ states. Thus we expect our reduced state to
be orthogonal to a separable pure state, e.g. $|00\rangle$.

Translated to our nearest neighbour state Eq.(\ref{gnnentanglement})
we need that e.g. ${\rm Tr}(v_1^\dagger v_1^\dagger\,\rho_B\, v_1
v_1)$ vanishes. Since $\rho_B$ has to be strictly positive
(otherwise we could reduce the dimension of ${\cal B}$) it follows
that $v_1$ has to be nilpotent, i.e. $v_1 v_1=\mathbf{0}$. Then the
nearest neighbour density matrix Eq.(\ref{gnnentanglement}) gets the
form
\begin{widetext}\label{rho12nilpotentA}
\begin{eqnarray} \rho_{12}&=&
\begin{pmatrix}
0&0&0&0\cr 0&{\rm Tr}(v_2^\dagger v_1^\dagger\;\rho_B\;v_1 v_2)&{\rm
Tr}(v_2^\dagger v_1^\dagger\;\rho_B\;v_2 v_1)&{\rm Tr}(v_2^\dagger
v_1^\dagger\;\rho_B\;v_2 v_2)\cr 0&{\rm Tr}(v_1^\dagger
v_2^\dagger\;\rho_B\;v_1 v_2)&{\rm Tr}(v_1^\dagger
v_2^\dagger\;\rho_B\;v_2 v_1)&{\rm Tr}(v_1^\dagger
v_2^\dagger\;\rho_B\;v_2 v_2)\cr 0&{\rm Tr}(v_2^\dagger
v_2^\dagger\;\rho_B\;v_1 v_2)&{\rm Tr}(v_2^\dagger
v_2^\dagger\;\rho_B\;v_2 v_1)&{\rm Tr}(v_2^\dagger
v_2^\dagger\;\rho_B\;v_2 v_2)\cr
\end{pmatrix}\nonumber\\
&=&
\begin{pmatrix}
0&0&0&0\cr 0&{\rm Tr}(v_1^\dagger \;\rho_B\;v_1)&{\rm
Tr}(v_2^\dagger v_1^\dagger\;\rho_B\;v_2 v_1)&{\rm
Tr}(v_1^\dagger\;\rho_B\;v_2)\cr 0&{\rm Tr}(v_1^\dagger
v_2^\dagger\;\rho_B\;v_1 v_2)&{\rm Tr}(v_1^\dagger
\;\rho_B\;v_1)&{\rm Tr}(v_1^\dagger\;\rho_B\;v_2)\cr 0&{\rm
Tr}(v_2^\dagger\;\rho_B\;v_1)&{\rm
Tr}(v_2^\dagger\;\rho_B\;v_1)&1-2\,{\rm Tr}(v_1^\dagger
\;\rho_B\;v_1)\cr
\end{pmatrix}
\nonumber\\
 &=&\left(\begin{array}{cccc}
0&0&0&0\\
0&A&B&C\\
0&B^*&A&C\\
0&C^*&C^*&1-2 A
\end{array}\right)%\nonumber\\
%&=&\frac{1}{4}\lbrace \mathbbm{1}_2\otimes \mathbbm{1}_2+(1-4
%A)\sigma_3\otimes\sigma_3+2
%B(\sigma_1\otimes\sigma_1+\sigma_2\otimes\sigma_2)+ (2
%A-1)(\sigma_3\otimes \mathbbm{1}_2+\mathbbm{1}_2\otimes\sigma_3)\nonumber\\
%& &+ 2 C
%(\sigma_1\otimes\mathbbm{1}_2+\mathbbm{1}_2\otimes\sigma_1+\sigma_1\otimes\sigma_3+\sigma_3\otimes\sigma_1)\rbrace
%\qquad\sigma_i\,=\,\text{Pauli matrices}\,.
\end{eqnarray}
\end{widetext}
This form of the density matrix is similar to the choice in
Ref.~\cite{WoottersEntangledChains} except that $C$ is required to be
equal zero there. It means that the state considered in the reference
is also invariant under local rotation of one qubit around the $x$--
and $y$--axis, a crucial assumption in the construction.  As we drop
this assumption, we can test another class of candidates within our
framework.

Let us now discuss the properties of a density matrix of the form
Eq.~(\ref{rho12nilpotentA}). Its eigenvalues are $\lbrace 0, A-|B|,
1-(A-|B|)\pm\sqrt{(3 A+|B|-1)^2+8 |C|^2}\rbrace$ and $A\in[0,1/2]$. First we
consider its purity as measured by ${\rm Tr}\rho_{12}^2$ which equals to $1-4
A+6 A^2+2 |B|^2+4 |C|^2$ and the purity for the reduced matrix, i.e. the
one--qubit state, equals to $Tr\rho_1^2=1-2 A+2 A^2+2 |C|^2$. Intuitively, we
expect the entanglement to increase for a density matrix becoming purer, while
the purity of the one--qubit state should decrease. However we will notice that
for the  two--qubit state the opposite is true.

As a measure of entanglement we use the concurrence ${\cal C}$, introduced by
Hill and Wootters~\cite{WoottersHillConcurrence,WoottersConcurrence}, which is
a monotonically increasing function of the entanglement of
formation~\cite{Benn}. The concurrence of a density matrix $\rho$ is given by
$\mathcal{C}(\rho)=\max\lbrace
0,\lambda_1-\lambda_2-\lambda_3-\lambda_4\rbrace$, where $\lambda_j$ are the
square roots of the eigenvalues in decreasing order of the matrix
$\rho\widetilde{\rho}$ where $\widetilde{\rho}=(\sigma_y\otimes\sigma_y) \rho^*
(\sigma_y\otimes\sigma_y)$ and $\rho^*$ denotes complex conjugation in the
standard basis. For the above form of density matrices the eigenvalues
$\sqrt{\rho\tilde\rho}=\{A+|B|, A-|B|, 0, 0\}$ are independent of $C$ and the
concurrence is simply ${\cal
  C}(\rho_{12})=2\,|B|$. Clearly, we have separability only for
vanishing $|B|$.

Another useful quantity is concurrence of
assistance~\cite{Entass,Cass} which is defined as the sum of
the square roots of the eigenvalues of $\rho\widetilde{\rho}$, i.e.
in our case its simply given by ${\cal C}_{\text{ass}}(\rho_{12})=2 A$.
This quantity characterizes the maximum entanglement of a selected
pair of qubits available on average when the rest of the system is
subjected to measurements.

In order to maximize nearest neighbour entanglement we have to
maximize the function
\begin{eqnarray}
{\cal C}(\rho_{12})=2\,\left|B\right|=2\, \left|{\rm Tr}(v_2^\dagger
v_1^\dagger\;\rho_B\;v_2 v_1)\right|\,.
\end{eqnarray}
There is an additional symmetry of the nearest neighbour state,
namely $\rho_{12}\rightarrow\rho_{12}^*$ does not change the amount
of entanglement. For dimension $b=2$  the only effect is to
introduce irrelevant phase factors to ${\rm Tr}(v_2^\dagger
v_1^\dagger\;\rho_B\;v_2 v_1)$. Though for higher dimensions $b$
this is not the case, we restrict ourselves to real generators as a
natural choice to reduce the set of parameters in the calculations.
We have checked complex extensions numerically in the region of the
obtained maxima where we always have found that it only reduces the
amount of entanglement.

In the next section we analyze the case of auxiliary matrices of
dimension $2$ where we can give analytical solutions and analyze the
generalization for higher dimension.

%XXXXX Remove XXXX\\
%-------------------------------------------------------\\
% Choice nilpotent-->form of matrix, general properties,
%entanglement of $1$ with $3$ not high because of nilpotent???
%
%
%Nilpotent: in 2x2 only matrix is ${{a,-a^2/b},{b,-a}}$ and nilpotent
%to power 3 (has same form)
%
%Proposition(v1 nilpotent): If $v_1^\dagger\rho
%v_2\not=(v_1^\dagger\rho v_2)^{\cal T}$ then $\rho_{A\otimes B}$
%entangled.
%
%Different view if $C$ goes to zero:
%\begin{eqnarray}
%\rho_{12}&=&\frac{A-|B|}{2}|\psi^-\rangle\langle\psi^-|+\frac{A+|B|}{2}|\psi^+\rangle\langle\psi^+|+(1-2 A)|11\rangle\langle 11|
%\nonumber\\
%&=&\frac{{\cal C}_{\text{ass}}-{\cal
%C}}{4}|\psi^-\rangle\langle\psi^-|+\frac{{\cal C}_{\text{ass}}+{\cal
%C}}{4}|\psi^+\rangle\langle\psi^+|+(1-{\cal
%C}_{\text{ass}})|11\rangle\langle 11|
%\end{eqnarray}
%We will see that the difference ${\cal C}_{\text{ass}}-{\cal C}$ minimizes
%and the sum ${\cal C}_{\text{ass}}+{\cal C}$ maximizes, see table
%\ref{table1}. (for 6x6 the difference is $\approx
%\sqrt{2}-\frac{3117}{2500}$)
%
%WOOTTERS Example for satisfying CKW ineuality?!
%-------------------------------------------------------\\

\subsection{Analytical results}

For dimension $b=2$ we can give analytical results and go through
the whole calculation in order to understand the procedure in more
detail and its generalization for higher dimensions. This will also
strengthen our assumption for the generators of the completely
unital map $\mathcal{E}$. First we investigate the set of solutions
$\rho_B$ for a nilpotent $v_1$ and its generalization for higher
dimensions. Then we discuss the implementation to the entanglement
to the nearest and next-nearest neighbours.

\subsubsection{Solutions for the auxiliary
density matrix $\rho_B$}

We have to solve the two conditions: unitality
Eq.~(\ref{condition1}) and translation invariance
Eq.~(\ref{condition2}). In $b=2$ dimensions the only nilpotent
matrices are the ladder operators where we choose without loss of
generality one of them with a weight $\cos\alpha_1$:
\begin{eqnarray}\label{nilpoten2x2}
v_1&=&\cos{\alpha_1} \left(\begin{array}{cc}
0&0\\
1&0
\end{array}\right)\, ,\nonumber\\
v_2&=&\left(\begin{array}{cc}
1&0\\
0&\sin{\alpha_1}
\end{array}\right).\left(\begin{array}{cc}
\cos{\phi_1}&\sin{\phi_1}\\
-\sin{\phi_1}&\cos{\phi_1}
\end{array}\right)\, ,
\end{eqnarray}
where $v_2$ is the most general (real) solution satisfying the
unitality condition in Eq. Eq.~(\ref{condition1}).

\begin{figure*}
\center{(a)\includegraphics[width=150pt, keepaspectratio=true]{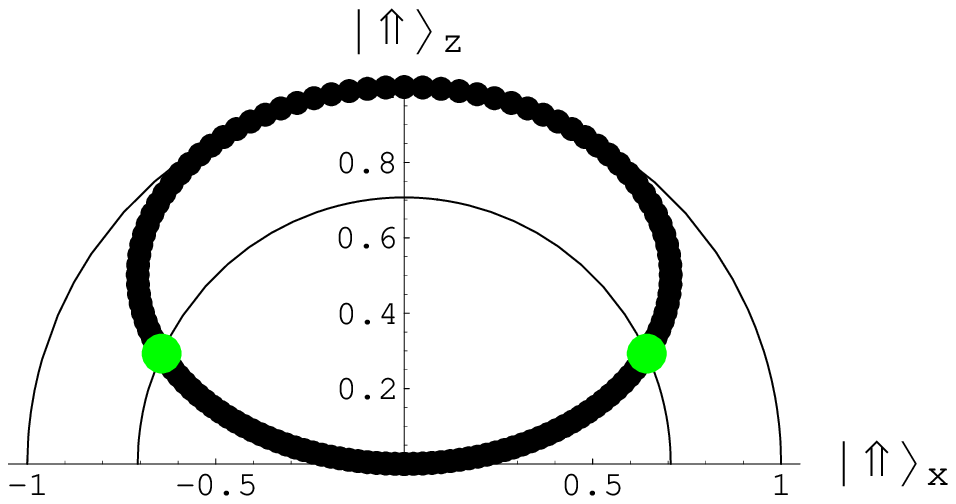}
%\includegraphics[width=200pt,
%keepaspectratio=true]{solutionrhoB.eps}
(b)\includegraphics[width=175pt, keepaspectratio=true]{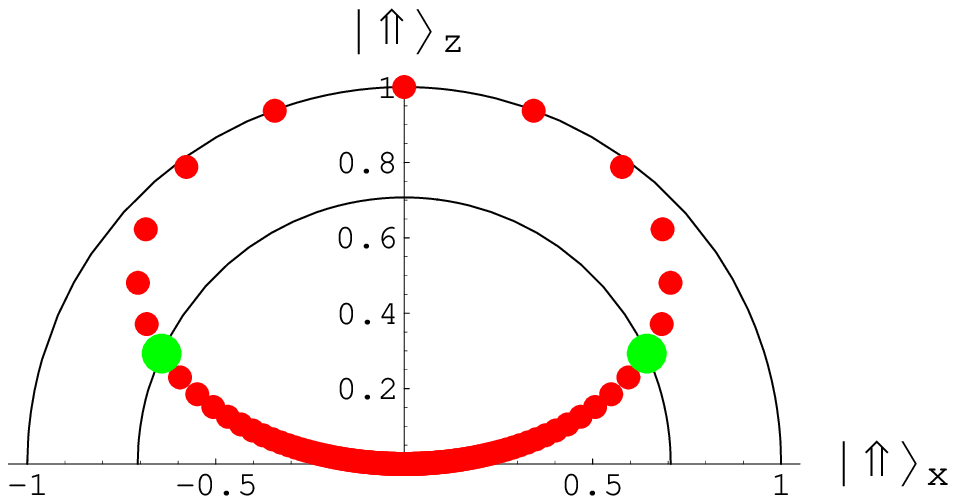}
(c)\includegraphics[width=175pt, keepaspectratio=true]{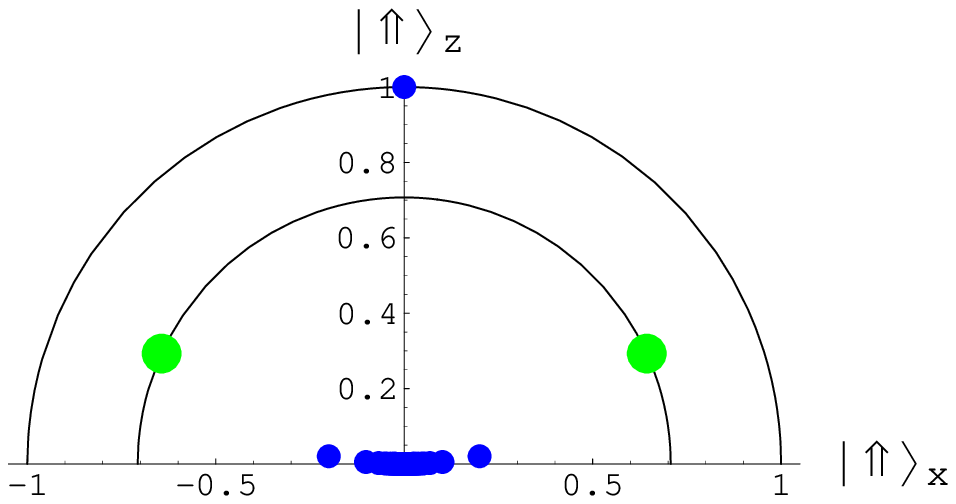}}
\caption{(Color online.) In the Figs.(a)-(c) the upper half of the $x$-plane of
the
  Bloch sphere is plotted. The possible $\rho_B$'s and the nilpotent
  $v_1$ satisfying Eq.~(\ref{nilpoten2x2}) form the ellipse given in
  Eq.~(\ref{ellipse}). The dots in Fig.(a)/(b)/(c) represent the
  solution for the Bloch vectors for
  $\alpha_1=0/\frac{\pi}{3}/0.47\pi$, where $\phi_1$ varies
  $\in[-\frac{\pi}{2},\frac{\pi}{2}]$ with a step size $\pi/128$. For
  $\alpha_1=0$, Fig.(a), the solutions cover homogenously the ellipse
  and for $\alpha_1\rightarrow \pi/2$ they concentrate at the tracial
  state, i.e. there is an decrease of the set of possible solutions.
  For $\phi=0$ the pure spin up $|\Uparrow\rangle$ is always a
  solution. The two big (green) dots are the solutions for which the
  nearest neighbour entanglement maximizes, this is when the Bloch
  vector $|\vec{n}^{(\text{Bloch})}|=1/\sqrt{2}$ which is also plotted
  (inner half circle). For higher dimensions $b$ the absolute value of
  the generalized Bloch vector $\vec{n}^{(\text{Bloch})}$ is always
  around $\approx 1/\sqrt{2}$, see Table \ref{table1}. Hence maximal
  entanglement between nearest neighbours is obtained if the ``rest of
  the chain'' is quite equally weighted in its purity and mixedness,
  measured by the squared length of the Bloch vector.}
\label{solutionofrho2x2}
\end{figure*}

Let us now discuss the solutions for the auxiliary $\rho_B$, conditions in
Eq.~(\ref{condition2}). The set of possible density matrices which are
invariant under the above chosen interaction is illustrated in
Fig.~\ref{solutionofrho2x2}, where we used the Bloch sphere representation.
Every one--qubit state can be decomposed into three Pauli matrices
\begin{eqnarray}
\rho_B &=&\frac{1}{2}(\mathbbm{1}_2+n_i^{(\text{Bloch})} \sigma^i), \qquad
    n_i^{(\text{Bloch})} \in \mathbbm{R}\,,\nonumber\\
    && \; \sum_i (n_i^{(\text{Bloch})})^2 = \left| \vec{n}^{\,(Bloch)} \right|^2 \leq 1 \,.
\end{eqnarray}
For $\left|\vec{n}_{(Bloch)}\right|^2 < 1$ the state is mixed
(corresponding to Tr$\,\rho^2 < 1$) whereas for
$\left|\vec{n}^{\,(Bloch)}\right|^2 = 1$ the state is pure
(Tr$\,\rho^2 = 1$). This real three dimensional vector
$\vec{n}^{(\text{Bloch})}$ is called the Bloch vector and thus the state
space of a qubit can be represented by a sphere, where the vectors
with $|\vec{n}^{\,(Bloch)}|=1$ are pure and cover the surface of the
sphere, inside the sphere we have all mixed states and the origin
represents the totally mixed state, i.e. the tracial state. As we
consider real generators of the interaction the $y$-component of
$\vec{n}^{\,(Bloch)}$ is zero and all possible one--qubit states are
represented by the area of a circle. And because of the specific
choice of the ladder operator only states in the upper half can
occur as solutions. It turns out that the solution is an ellipse in
this Bloch's sphere, i.e. the following equation holds for all
$\alpha_1, \phi_1$
\begin{eqnarray}\label{ellipse}
\frac{(n_1^{(\text{Bloch})}-0)^2}{(\frac{1}{\sqrt{2}})^2}+\frac{(n_3^{(\text{Bloch})}-\frac{1}{2})^2}{(\frac{1}{2})^2}=1\,
,
\end{eqnarray}
where the Bloch components are $n_1^{(\text{Bloch})}=2[\rho_B]_{12}$ and
$n_3^{(\text{Bloch})}=2[\rho_B]_{11}-1$.

For $b=3$ the auxiliary density matrix $\rho_B$ is described by a qutrit state
which can be decomposed analogously to the qubit case into Gell-Mann matrices
$\lambda^1,\dots ,\lambda^8$ (consult the Appendix for their definitions)
\begin{eqnarray}
    \rho_B &=& \frac{1}{3} \left( \mathbbm{1} + \sqrt{3} \,n_i^{(\text{Bloch})} \,\lambda^i \right), \qquad
    n^{(\text{Bloch})}_i \in \mathbbm{R}\,,\nonumber\\
    && \; \sum_i (n_i^{(\text{Bloch})})^2 = \left| \vec{n}^{\,(Bloch)} \right|^2 \leq 1 \;.
\end{eqnarray}
where the Bloch vector is a $8$ dimensional real vector with similar properties
as in the qubit case. Notice that the positivity of the state does not hold for
all vectors in the $8$ dimensional sphere. The Gell-Mann matrices satisfy the
similar relations as the the Pauli matrices, i.e. $\rm{Tr}\,\lambda^i = 0, \;
\rm{Tr}\,\lambda^i \lambda^j = 2\, \delta^{ij}$. If we choose an analogous
nilpotent generator
\begin{eqnarray}\label{nilpoten3x3}
v_1=\cos{\alpha_1} \left(\begin{array}{ccc}
0&0&0\\
1&0&0\\
0&0&0
\end{array}\right),\; v_2=diag(1,\sin\alpha_1,1).R
\end{eqnarray}
where $R$ is a $3\times 3$ (real) orthogonal matrix which we can
build up with three angles each representing a rotation in the
two-dimensional subspace. It turns out that the solution of the
Bloch vectors for $\rho_B$ again is an ellipse, more precisely a $5$
axial ellipsoid if we fix the three angles and vary only $\alpha_1$.
Varying one angle we obtain again an ellipsoid but with a different
center and semi-axes.

For the density matrix $\rho_B$ which maximizes nearest neighbour
entanglement the length of the Bloch vector in $b=2$ is
$|\vec{n}^{\,(Bloch)}|=\frac{1}{\sqrt{2}}$. It turns out that for
higher dimensions $b$ the length of the generalized Bloch vector is
always around $\frac{1}{\sqrt{2}}$, see Table~\ref{table1}. The
``purity'' of the state of the rest of the chain $\rho_B$ measured
by the squared length of the Bloch vector
$|\vec{n}^{\,(Bloch)}|^2=\frac{b Tr\rho_B^2-1}{b-1}$ seems to be
quite constant when nearest neighbour entanglement is optimized.
%\\
%XXXXXXXXXXXXXXXXXXXXXXXXXXXXXXXXXXXXXXXXXXXXXXxx\\
%Generally for qu$d$its we can define the same decomposition. Is
%there a vector to which it converges, one angle? Not with$\lbrace
%1,1,\dots,1\rbrace$. For nearest neighbour entanglement ${\rm Tr}
%(v_1^\dagger \rho v_1)\approx 0.30$. length constant, is there a
%general angle where it
%converge to?XXXXXXXXX\\

\subsubsection{Properties of the nearest neighbour density matrix
$\rho_{12}$}

\begin{figure*}
\center{
\includegraphics[width=200pt, keepaspectratio=true]{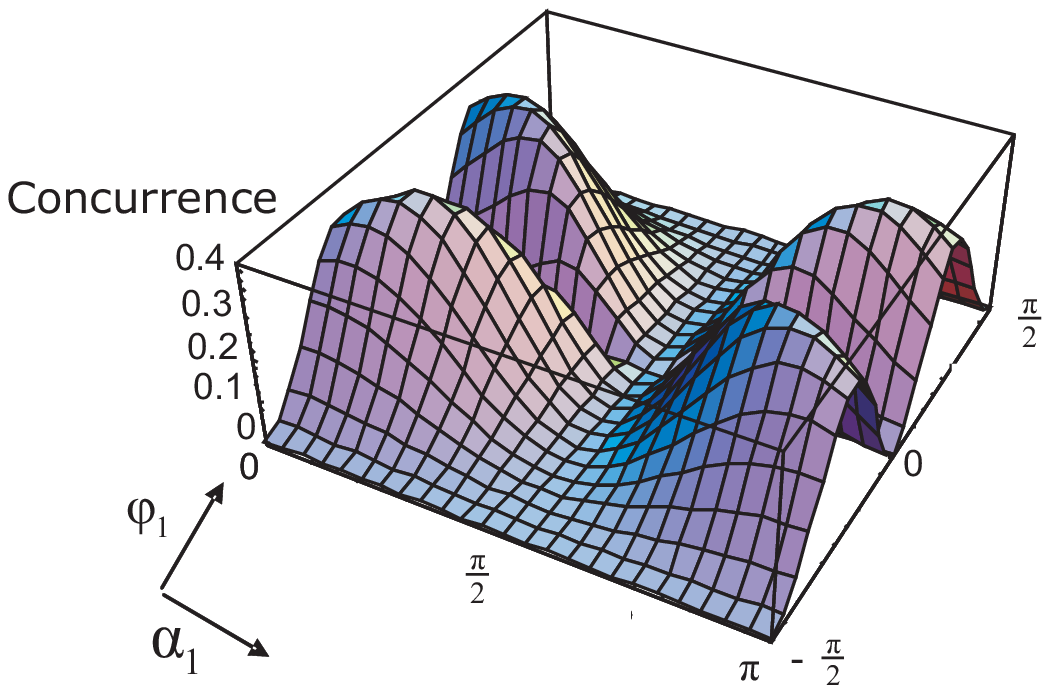}\hspace{1cm}\includegraphics[width=200pt, keepaspectratio=true]{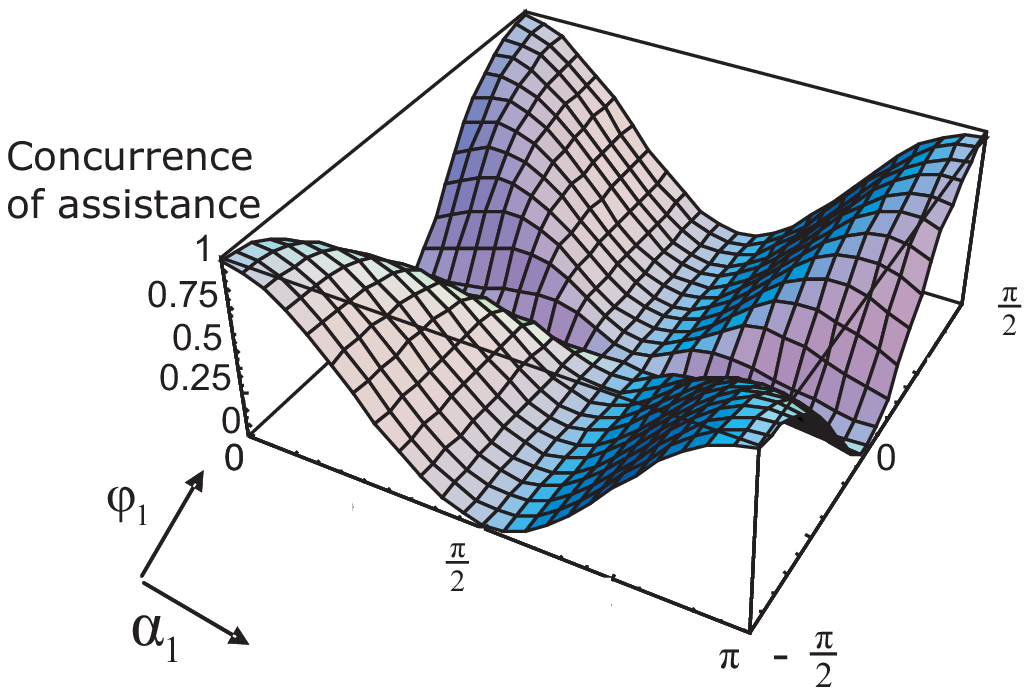}
\caption{(Color online.) In Fig. (a) we have plotted the concurrence
  $\mathcal{C}_{12}(\rho_{12})$ of the nearest neighbour entanglement
  in Eq.~(\ref{concurrence}) for the parameters $\alpha_1\in[0,\pi]$
  and $\phi_1\in[-\frac{\pi}{2},\frac{\pi}{2}]$. The concurrence
  function is invariant under a shift of $\alpha_1$ with $\pi/2$ and
  $\phi\rightarrow -\phi$, thus we obtain four local maxima.\\
  In Fig. (b) the concurrence of assistance
  $\mathcal{C}_{\text{ass}}(\rho_{12})$ in Eq.~(\ref{assistance}) is
  plotted. All possible values can be obtained.}All plotted
quantities are dimensionless.}\label{C2x2}
\end{figure*}

Let us now return to the original question, i.e. to the function that maximizes
nearest neighbour entanglement. We have noticed in
Sec.\ref{generaldiscussionv1} that due to the nilpotent choice of $v_1$ the
trace of $\rho\tilde\rho$ gives only two non--vanishing eigenvalues, i.e.
$\lbrace A+|B|,A-|B|\rbrace$. Concurrence of nearest entanglement for dimension
$b=2$ is therefore
\begin{eqnarray}\label{concurrence}
&{\cal C}(\rho_{12})\;=\; 2\, {\rm Tr} (v_2^\dagger v_1^\dagger\,
\rho_B\,
v_2 v_1)= 2\, \left|B\right|&\nonumber\\
&=\frac{\cos^2\alpha_1\,\left( 1 + \sin\alpha_1 \right)
\,\sin^2\phi_1\,\cos^2\phi_1}
 {\cos^2\alpha_1\,\cos^2\phi_1\,\left( -1 +
\sin\alpha_1 \right) -
    2\,\left( 1 + \sin\alpha_1 \right) \,\sin^2\phi_1}\,,&
\end{eqnarray}
and concurrence of assistance
\begin{eqnarray}\label{assistance}
&{\cal C}_{\text{ass}}(\rho_{12})\;=\;2\,{\rm Tr}(v_1^\dagger\, \rho_B\,
v_1)=2\, A &\nonumber\\
&= \frac{\cos^2\alpha_1\,\left( 1 + \sin\alpha_1 \right)
\,\sin^2\phi_1} {\cos^2\alpha_1\,\cos^2\phi_1\,\left( -1 +
\sin\alpha_1 \right) -
    2\,\left( 1 + \sin\alpha_1 \right) \,\sin^2\phi_1}\,.&
\end{eqnarray}
We have plotted both functions in Figure~\ref{C2x2}. One notices that while one
can obtain for the concurrence of assistance all possible values, i.e. ${\cal
C}_{\text{ass}}(\rho_{12})\in[0,1]$, concurrence of nearest neighbour entanglement has
a maximum value of ${\cal C}_{\text{max}}(\rho_{12})=\sqrt{2}-1=0.41421$ for
$\alpha_1=0.427079, \phi_1=0.571859$. This gives a concurrence of assistance of
${\cal C}_{\text{ass}}(\rho_{12})=0.585787$.

Further one notice that concurrences and concurrence of assistance are only
equal for $\alpha_1$ or $\phi_1$ equal zero.

Armed with this analytical experience we proceed to the numerical procedure and
present the results for increasing dimensionality $b$ of the auxiliary system.

\section{Numerical optimization of nearest neighbour
entanglement}\label{numericaloptimization}

First we discuss the parametrization and our different strategies to
numerically maximize nearest neighbour entanglement, then we discuss
the results of the maximum in different dimensions $b$. Then we
proceed with a discussion of the properties of such a chain
maximizing nearest neighbour entanglement.

\subsection{Parametrization for dimension $b$}

The triple $({\cal B}, \mathbb{E},\rho_B)$ defining the finitely correlated
state is obtained by a finite number of parameters. We choose $\mathcal{B}={\bf
M}_b$, and carry out calculations for different dimensionalities $b$.  The
completely positive map $\mathbb{E}$ is described by the two matrices $v_1$ and
$v_2$ which are $b\times b$ matrices and have to satisfy the conditions
Eq.~(\ref{condition1}, \ref{condition2}).

We choose $v_1$ to be a nilpotent operator as argued in the previous
section, with a matrix of the form
\begin{equation}
  \label{eq:v1choice}
  v_1=
  \begin{pmatrix}
    0             & &              & \cr
    \cos(\alpha_1)&0&              & \cr
                  &0&0             & \cr
                  & &\cos(\alpha_2)&0 \cr
                  & &              &0&0\cr
                  & &              &&\ddots&\ddots
\end{pmatrix},
\end{equation}
described by $[b/2]$ real parameters, $[\ldots]$ denoting the
integer part.  Though the parametrization is periodic with a
periodicity of $2\pi$ in each parameter, the parameter values are
unconstrained, which is an important simplification in the case of
numerical optimization. In order to satisfy the unitality condition
in Eq.~\eqref{condition1}, we set
\begin{widetext}
\begin{equation}
  \label{eq:v2choice}
    v_2=\diag (1, \sin(\alpha_1), 1, \sin(\alpha_2),\ldots)\, R,
\end{equation}
where $R$ is an arbitrary $b\times b$ unitary matrix. However,
according to our numerical experience for up to $b=6$ dimensions
supports the conjecture that it is enough to consider real
orthogonal matrices as $R$. The introduction of general unitary
$R$-s did not lead to the increase of the maximal nearest neighbour
concurrence. Thus we build up the generic $R\in SO(b)$ from
rotations in two-dimensional subspaces, yielding the following
(periodic, unconstrained) parametrization of $R$, with $b(b-1)/2$
parameters~\cite{Murnaghan}:
\begin{eqnarray}
  \label{eq:Rchoice}
  R&=&
    \begin{pmatrix}
    \cos(\phi_1) &\sin(\phi_1) & 0 &\hdots\cr
    -\sin(\phi_1)&\cos(\phi_1) & 0 &\hdots\cr
        0        &0         & 1 &\hdots \cr
        \vdots        &\vdots         & \vdots &\ddots
  \end{pmatrix}
\begin{pmatrix}
    \cos(\phi_2) &0 &\sin(\phi_2) &\hdots\cr
    0             &1 & 0            &\hdots \cr
    -\sin(\phi_2)&0 &\cos(\phi_2) &\hdots\cr
    \vdots        &\vdots        &\vdots & \ddots
  \end{pmatrix}
  \ldots
\begin{pmatrix}
    \cos(\phi_{b-1}) & \hdots &\sin(\phi_{b-1}) \cr
     0             & \hdots &  0          \cr
    \vdots &  & \vdots&            \cr
    -\sin(\phi_{b-1})&\hdots &\cos(\phi_{b-1})
  \end{pmatrix}
\nonumber \\ & &\times
  \begin{pmatrix}
    1 & 0 &0 &\hdots   \cr
    0 &\cos(\phi_{b})& \sin(\phi_{b})  &  \hdots          \cr
    0 &-\sin(\phi_{b})&\cos(\phi_{b})  & \hdots\cr
    \vdots &  & \vdots&
  \end{pmatrix}
 \ldots
  \begin{pmatrix}
    \ddots &  & \vdots &\vdots   \cr
     \hdots &      1   &      0          &  0\cr
    \hdots & 0 &\cos(\phi_{b(b-1)/2})& \sin(\phi_{b(b-1)/2})  \cr
    \hdots & 0 &-\sin(\phi_{b(b-1)/2})&\cos(\phi_{b(b-1)/2})
  \end{pmatrix}\, .
\end{eqnarray}
\end{widetext}
Thus given the dimensionality $b$, and a set of parameters
$\underline{\alpha}=\alpha_1\ldots \alpha_{[b/2]}$,
$\underline{\phi}=\phi_1\ldots \alpha_{b(b-1)/2}$, we can readily
evaluate $v_1(\underline{\alpha})$ and
$v_2(\underline{\alpha},\underline{\phi})$. Having these matrices at
hand, we can calculate numerically $\rho$ from the translational
invariance condition in Eq.~\eqref{condition2}. This can be done by
noticing that Eq.~\eqref{condition2} is linear in the matrix
elements of $\rho$, thus we have to calculate the nullspace of the
linear mapping
\begin{equation}
  \label{eq:lmap}
  L(\rho)=\sum_{j=1}^2v_j^\dagger\ \rho\ v_j-\rho,
\end{equation}
in the linear space of $b\times b$ matrices, in which all the
vectors $\rho$ are suitable for our aims.  We have found that for
all the parameter settings arising in our optimization procedure
$\rank L=b-1$ holds within the numerical precision, therefore the
nullspace is one-dimensional. Hence for a fixed $b$ and parameters
$\underline{\alpha},\underline{\phi}$, in addition to
$v_1(\underline{\alpha})$ and
$v_2(\underline{\alpha},\underline{\phi})$, we obtain a unique
$\rho(\underline{\alpha},\underline{\phi})$. As a numerical check we
verified that the solution is Hermitian positive semidefinite in all
cases which have occurred.

Performing the above calculations we can compute the
nearest-neighbour density matrix. From this density matrix we can
evaluate the concurrence.

Thus for a fixed dimensionality $b$, we have a function
$\mathcal{C}(\underline{\alpha},\underline{\phi})$ which we can
numerically evaluate. This is the subject of an unconstrained
numerical maximization in terms of its parameters.  Unfortunately,
it is not a convex function, thus there is no warranty to find a
global maximum numerically. In addition the function might be not
differentiable at certain points due to the properties of
concurrence. Therefore we chose the simulated annealing method,
which is known to be effective for mildly nonconvex and
non-differentiable function. We have used the routines available in
the MINTOOLKIT~\cite{Creel04} package of GNU Octave
software~\cite{Octavemanual}.  First we have searched for the
maximum using the {\tt samin} routine, a simulated annealing code
based on the implementation by Goffe~\cite{Goffe96}. We have set the
control parameters of the routine to $nt=20$, $ns=10$ $rt=0.75$,
$neps=5$ and $eps=10^{-10}$ (consult the
documentation~\cite{Creel04} of the routine for their exact
meaning). The routine showed a normal convergence in each case. Then
the so-obtained maxima were used as an initial condition for a
conjugate gradient search {\tt bfgsmin}, with numerical gradient. We
have found that the function is indeed differentiable around this
maximum. The conjugate gradient search showed a strong convergence.
The so obtainable final result is somewhat more accurate than the
one obtained directly from simulated annealing.  As a result of
these procedure, we have obtained the parameter sets
$\underline{\alpha},\underline{\phi}$ for which the
nearest-neighbour concurrence
$\mathcal{C}(\underline{\alpha},\underline{\phi})$ has a maximum
value. Though this procedure does not give a full warranty for
finding the global maximum, it is very likely that the obtained
maxima are indeed global.

\subsection{Numerical results of the maximum nearest neighbour entanglement}

We have summarized the results of the above described optimization
procedure in Table~\ref{table1}. In case of entangled chains it is
conjectured~\cite{WoottersEntangledChains} that the maximum value of
nearest neighbour entanglement as measured by concurrence is
${\mathcal{C}_W}=0.434467$.  As it is apparent from the results in
Table~\ref{table1}, in our framework we can obtain a state which
almost reaches this upper bound. Thus the translationally invariant
finitely correlated chains can approach the state of an entangled
chain with maximal bipartite entanglement quite fast. This is our
main result. The approximation improves with the increasing
dimensionality $b$ of the auxiliary Hilbert-space $\mathcal{B}$.

In addition, with accidental conjugate gradient searches we could
obtain local maxima $\mathcal{C}=0.43406$ for $b=8$ and
$\mathcal{C}=0.434095$ for $b=9$. These both constitute about $0.09\%$
relative difference from ${\mathcal{C}_W}=0.434467$.

We have to remark however, that as the maximum value of
${\mathcal{C}_W}$ is a conjecture, too, and we can neither fully
warrant the global maximum, nor check the $b\to \infty$ case in the
numerical framework, we cannot exclude the possibility to go beyond
${\mathcal{C}_W}$. Nevertheless we can prove explicitly that the
bound can indeed be obtained in the framework of FCS, even under
several restrictions.

Further we have checked numerically the possibility of using unitary
instead of orthogonal matrices for $R$, and also the application of
a more general, non-nilpotent $v_1$ by adding certain elements to
its upper diagonal. We have found for $b$ up to $6$ that this does
not improve the obtained maximal concurrence. In addition, the so
arising $v_1$ was always nilpotent, with numerically the same matrix
elements as in Eq.~\eqref{eq:v1choice}, though eventually ordered in
a different form in the matrix. This supports the assumptions we
have made as well as those in Ref.~\cite{WoottersEntangledChains}.

\begin{figure}
\center{
\includegraphics[width=250pt, keepaspectratio=true]{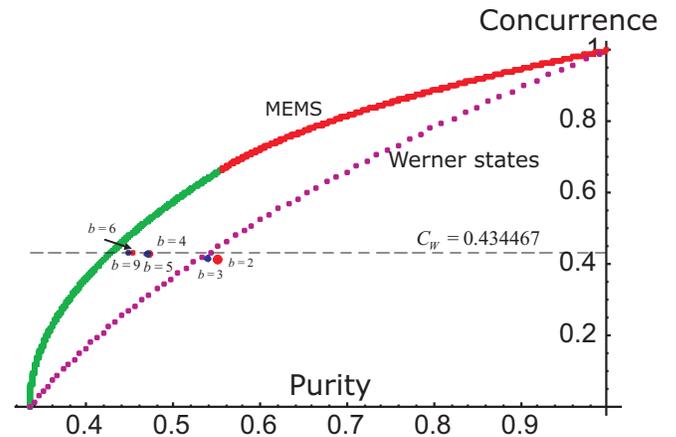}}
\caption{(Color online) Here concurrence versus purity measured by ${\rm
Tr}\rho^2$ is plotted. The points below the curve represent all possible
two-qubit density matrices. The curve itself are the maximally entangled mixed
states (MEMS), first introduced in Ref.~\cite{IshizakaHiroshima}. The MEMS are
defined up to local unitary transformations by $\rho^{\text{MEMS}}=
(\frac{1}{3}+\frac{q}{2})
|\psi^+\rangle\langle\psi^+|+(\frac{1}{3}-\frac{q}{2})
|\psi^-\rangle\langle\psi^-|+\frac{1}{3} |11\rangle\langle 11|$ for
$q\in[0,\frac{2}{3}]$ (green) and $\rho^{\text{MEMS}}= q
|\psi^+\rangle\langle\psi^+|+(1-q) |11\rangle\langle 11|$ for
$q\in[\frac{2}{3},1]$ (red). The dotted curve below (pink) are the density
matrices of the isotropic states, the so--called Werner states
$\rho_{\text{Werner}}=
\frac{1-p}{4}\mathbbm{1}\otimes\mathbbm{1}+p\,|\psi^+\rangle\langle\psi^+|$.
The horizontal dashed line represents all states having a concurrence of ${\cal
C}_\text{W}=0.434467$. The dots are the nearest neighbour density matrices
maximizing nearest neighbour entanglement where from left to right the
dimensionality $b$ of the auxiliary system increases. One notices that the
solutions seem to approach the bound of the realizable states, in particular
the MEMS states. Further note that the picture of entanglement measure versus
mixedness strongly depends on the chosen measure of entanglement and mixedness
and that different entanglement measures define different ordering of states,
see Ref.~\cite{WeiKwiatVerstraete,ZimanBuzek}.}\label{CversusP}
\end{figure}

As we explicitly can calculate elements of the nearest neighbour state (see
Table \ref{table2}), we can ask which final state is approached for enlarging
the dimensionality $b$ of the auxiliary system. For this we plotted the
obtained nearest neighbour density matrices maximizing entanglement in a
concurrence versus purity diagram (see Fig.~\ref{CversusP}). For dimension
$b=9$ the relative difference of concurrence and purity of the nearest
neighbour state maximizing entanglement and the maximally entangled mixed state
(MEMS) Ref.~\cite{IshizakaHiroshima}, in our case
$(\frac{1}{3}+\frac{1}{2}{\cal
C}_\text{W})|\psi^+\rangle\langle\psi^+|+(\frac{1}{3}-\frac{1}{2}{\cal
C}_\text{W})|\psi^-\rangle\langle\psi^-|+\frac{1}{3}|11\rangle\langle11|$, is
concerning concurrence $0.09\%$ and concerning purity $1.6\%$.

\begin{center}
\begin{table*}[htbp]
  \begin{tabular}{|l|c|c|c|c|c|c|c|}
    \hline
    Dimensionality $b$ & 2 & 3 & 4 & 5 & 6 & 7 \\
    \hline
    \hline
    Nearest-neighbour concurrence
    & 0.41421&0.41825 & 0.43200& 0.43247 & 0.43336&  0.43381 \\
    \hline
    Relative difference  & & & & & & \\
    from Wootters' bound (\%) &
                      4.66 & 3.73 & 0.57 & 0.46 & 0.25 & 0.15 \\
    \hline
    \hline
$\alpha_1$  & 0.427079 & 3.27378 & 0.252679& 6.345324 & 3.84312 &  2.71122 \\
    \hline
$\alpha_2$  &          &         &  2.888910& 0.269592& 0.10177 &  3.14860 \\
    \hline
$\alpha_3$  &          &         &         &          & 3.10541 &  3.29590 \\
    \hline
    \hline
$\phi_1$  & 0.571859  & 3.14062 &  0.062823& 6.22996 & 5.88873& 6.27750  \\
    \hline
$\phi_2$  &          &  0.56623 & 5.504548 & 2.351162& 6.10731& 2.50188  \\
    \hline
$\phi_3$  &          &  4.17472 & 5.892460&  2.713085& 1.48352& 3.33956  \\
    \hline
$\phi_4$  &          &         & 0.805037 & 0.047930 & 4.71882& 6.25125  \\
    \hline
$\phi_5$  &          &         &  0.272233&  5.137121& 1.38430& 5.62825   \\
    \hline
$\phi_6$  &          &         & 0.741237 & 0.417055 & 0.79196& 3.76442   \\
    \hline
$\phi_7$  &          &         &         & 5.628356  & 4.81583& 1.09039    \\
    \hline
$\phi_8$  &          &         &         &  1.759880 & 2.01345&  3.43100   \\
    \hline
$\phi_9$  &          &         &         &  5.728579 & 0.306965&  3.23516  \\
    \hline
$\phi_{10}$  &          &         &         &  1.193187 & 5.68444& 2.87925   \\
    \hline
$\phi_{11}$  &          &         &         &          & 6.03621& 4.95371    \\
    \hline
$\phi_{12}$  &          &         &         &          & 0.65283& 0.28542    \\
    \hline
$\phi_{13}$  &          &         &         &          & 5.67111&  1.87790   \\
    \hline
$\phi_{14}$  &          &         &         &          & 2.06680&  5.46657   \\
    \hline
$\phi_{15}$  &          &         &         &          & 1.78624& 1.14039    \\
    \hline
$\phi_{16}$  &          &         &         &          &         & 4.75900   \\
    \hline
$\phi_{17}$  &          &         &         &          &         &  2.68202  \\
    \hline
$\phi_{18}$  &          &         &         &          &         &  3.51887  \\
    \hline
$\phi_{19}$  &          &         &         &          &         &  5.54982  \\
    \hline
$\phi_{20}$  &          &         &         &          &         &   4.35086 \\
    \hline
$\phi_{21}$  &          &         &         &          &         &  0.478595 \\
    \hline
  \end{tabular}
  \caption{Optimal nearest neighbour concurrences found in FCS, and the corresponding parameters}
  \label{table1}
\end{table*}
\end{center}

\section{Properties of the translational chain maximizing nearest entanglement}
\label{sec:properties}

Let us investigate the question which properties an infinity translational
chain has which maximizes nearest neighbour entanglement. Is the obtained chain
like an ordinary bicycle chain (or Markov chain), whose links are only
connected to two neighbouring links but not to the next and next--next
neighbouring sites as our intuition may suggest?

Lets consider the state of three qubits in a line
\begin{eqnarray}
\rho_{[1,3]}&=&\sum |s_1 s_2 s_3\rangle\langle t_1 t_2 t_3|\;
Tr(v_{s_1}^\dagger v_{s_2}^\dagger v_{s_3}^\dagger\; \rho_B\;
v_{t_1} v_{t_2} v_{t_3})\nonumber\\
\end{eqnarray} and tracing over
qubit $2$ gives the density matrix for next nearest neighbours
\begin{eqnarray}
\rho_{13}&=&Tr_2(\rho_{[1,3]})\nonumber\\
&=&\sum |s_1 s_3\rangle\langle t_1 t_3| \left\{Tr(v_{s_1}^\dagger v_{1}^\dagger
v_{s_3}^\dagger \rho_B\; v_{t_1} v_{1}
v_{t_3})\right. \nonumber\\
&&\left. \qquad\qquad\qquad +Tr(v_{s_1}^\dagger v_{2}^\dagger v_{s_3}^\dagger
\rho_B\; v_{t_1} v_{2} v_{t_3})\right\}
\end{eqnarray} If $v_1$ is nilpotent, this means that the first term
of the the last equation vanishes (except of the last one but this has just to
do with normalization). Thus it reduces the space dramatically in which we can
vary. We checked that this next nearest neighbour entanglement for the
parameters maximizing the nearest entanglement and found that it is zero for
all auxiliary dimensions $b$. Thus we conclude that such a chain is an ``ordinary bicycle chain'': only two neighbouring sites are linked via entanglement. Thisalso supports our assumption for a nilpotent generator because entanglement
concentrates.

\begin{figure}
\center{
\includegraphics[width=200pt, keepaspectratio=true]{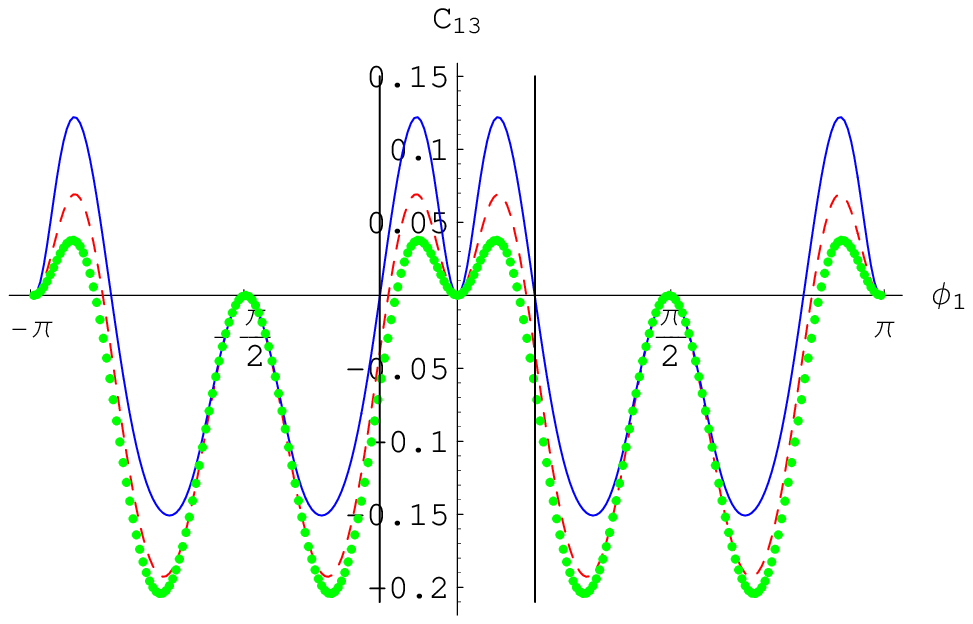}\hspace{3cm}
\includegraphics[width=200pt, keepaspectratio=true]{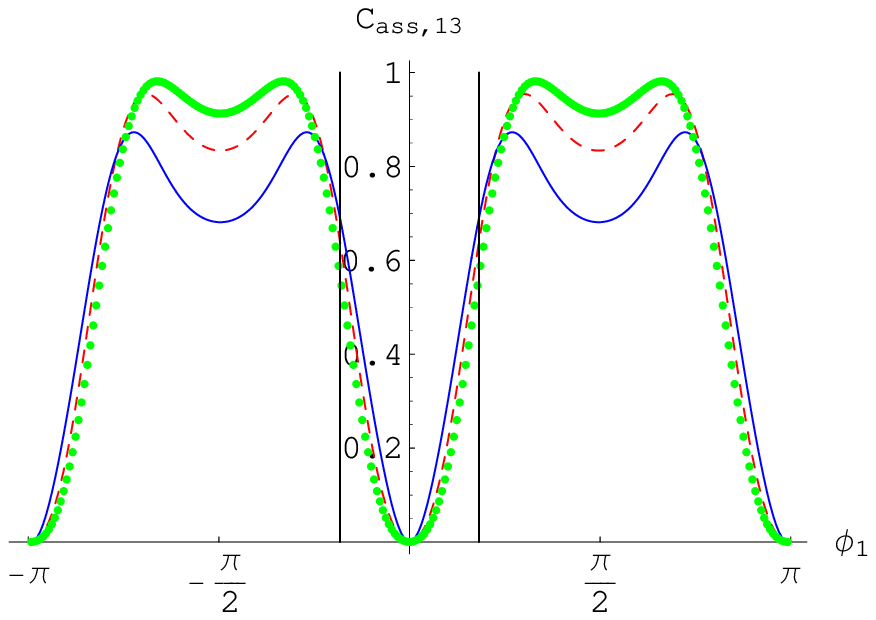}
\caption{(Color online) (a) Here we plot $\mathcal{C}(\rho_{13})$ for
$\alpha_1=0.6 \text{ (blue)},0.42 \text{ (red, dashed)}, 0.3 \text{ (green,
dotted)}$; the line is the maximizing $\phi_1=0.571859$ and the abscissa is
$\phi_1\in[-\frac{\pi}{2},\frac{\pi}{2}]$. Note that we plotted
$\lambda_1-\lambda_2$ while the concurrence is the maximum of
$\{|\lambda_1-\lambda_2|,0\}$. (b) Here the concurrence of assistance
$\mathcal{C}_{\text{ass}}=\rho_{13}$ for $\alpha_1=0.6 \text{ (blue)},0.42 \text{
(red, dashed)}, 0.3 \text{ (green, dotted)}$ is plotted; abscissa
$\phi_1\in[-\frac{\pi}{2},\frac{\pi}{2}]$. One notices when concurrence
increases, concurrence of assistance decreases and vice versa.} \label{c13}}
\end{figure}

In Fig.~\ref{c13} we plotted the concurrence and assisted concurrence for the
dimension $b=2$ for different parameters. One sees that we can generally find
parameters for which next nearest entanglement is nonzero, we checked the
maximum possible concurrence available for our choice of generators, which is
below the one of nearest neighbour entanglement ($C_{13}=0.169470$ for
$\alpha_1=0.88563$ and $\phi_1=0.25066$ and a nearest neighbour concurrence of
$C_{12}=0.270660$). When concurrence of next nearest neighbour entanglement
increases then concurrence of assistance of next nearest neighbour entanglement
decreases, as it is plotted in Fig.\ref{c13}~(b).

%Checked concurrence for 3x3 which is zero and Cass is $0.650769$.
%For 4x4 we get also concurrence zero and Cass increases $0.76918$.
%For 2x2 concurrence zero and Cass is $0.634747$. For 6x6 we have
%Cass $0.792718$. For 9x9 we have Cass $0.799776$.

As the state of the whole chain is pure, we expect that purity has
to increase considering the states of more and more qubits of the
chain. We checked that for all considered cases, for example for
$b=6$ $Tr \rho_{12}^2<Tr\rho_{123}^2$ is $0.452911<0.461722$ and for
$b=9$ it changes to $0.447191<0.455342$, i.e. far away from $1$.
Increasing dimensionality $b$ increases entanglement and also
concentrates the correlation.

In Table~\ref{table2} the relevant quantities are summarized. Notice that all
quantities behave monotonically with increasing dimensionality $b$, except for
the length of the Bloch vector. The fact that concurrence of assistance is
bigger than concurrence suggest the presence of multipartite entanglement in
the system.

%DISCUSSION of Table 2!!

\begin{center}
\begin{table*}
\begin{tabular}{|l||c|c|c|c|c|c|}
\hline
Dimension b& 2 & 3 & 4 & 5 & 6 & 9\\
\hline\hline ${\cal C}_{\text{max}}
(\rho_{12})$&$\;0.414214\;$&$\;0.41825\;$&$\;0.432000\;$&$\;0.432471\;$&$\;0.433791\;$&$\;0.434095\;$\\
 \hline ${\cal C}_{\text{ass}}(\rho_{12})$&$0.585787$&$0.587251$&$0.600000$&$0.600131$&$0.601204$&$0.601442$\\
 \hline
\hline $\mathrm{A}$& 0.292893 & 0.293626 & 0.300000 & 0.300066 &0.300602&0.300721\\
\hline $\mathrm{B}$& 0.207107 & 0.209126 & 0.216000 & 0.216236 &0.216895&0.217048\\
\hline $\mathrm{C}$& 0.174155 & 0.164125 & 0.097378 & 0.0925458 &-0.069033&-0.0575684\\
\hline\hline
Purity of ${\rm Tr} \rho_{12}^2$&0.550252&0.538009 &0.471242  & 0.467748 &0.452911&0.447191\\
\hline \hline
Purity of ${\rm Tr} \rho_{1}^2$&0.646446& 0.639055& 0.598965& 0.597077 &0.58905&0.586052\\
\hline\hline $|\vec{n}^{(\text{Bloch})}|^2=\frac{b {\rm Tr}\rho_B^2-1}{b-1}$ &0.5&0.607465& 0.536326 &0.554368& 0.519502&0.521177\\
\hline
\end{tabular}\caption{Properties for different dimension $b$}\label{table2}
\end{table*}
\end{center}

\section{Summary and conclusion}
\label{sect:summary}

We have addressed the optimization of entanglement between nearest neighbours
in an infinite translational invariant chain in an increasing set of states.
Using finitely correlated states with a recursive structure and certain
reasonable assumptions we have shown that the nearest neighbour entanglement
almost reaches its conjectured upper bound $C_{\text{W}}=0.434467$. Our
approach allows for an explicit calculation of the elements of the nearest
neighbour density matrix maximizing entanglement. We show that in a concurrence
versus purity (measured by $Tr\rho^2$) diagram the obtained nearest neighbour
state seems to approach the maximally entangled mixed states (MEMS) which bound
the realizable bipartite states in this diagram.

The approach we have adopted has the same roots as the DMRG methods. However,
instead of using a matrix product form for a state of a finite set of qubits
the formalism used provides an exact description of the infinite chain. The
accuracy of the approximation increases with the dimensionality $b$ of the
auxiliary Hilbert space.

For dimension $b=2$ and $b=3$ we have given a detailed analytical
treatment of the problem while in higher dimensions we rely on
numerical calculations. Apart from the investigation of nearest
neighbour entanglement we have evaluated other properties of the
chain.  These results support the qualitative expectations that due to
the monogamy of entanglement the increase in the nearest neighbour
entanglement leads to the decrease in the longer distance quantum
correlation. Not only concurrence but also the whole nearest neighbour
density matrix tends to reach a given fixed value. This is, however,
not the case with the state of the whole system. The difference of the
concurrence and the concurrence of assistance suggest that some kind
of multipartite entanglement is also present.

The variational technique utilized in our work may be applicable in
other similar physical problems.

\acknowledgments The authors wants to thank the ``non--local'' seminar
Vienna-Bratislava. B.C. Hiesmayr wants to further acknowledge the
EU-project EURIDICE HPRN-CT-2002-00311. M.Koniorczyk acknowledges the
support of the Hungarian National Grant Agency (OTKA) under the
contracts Nos. T043287 and T049234, the hospitality of Prof. V. Bu\v
zek in Bratislava during the first period of this project, and the
support of the Marie Curie RTN network CONQUEST.

\appendix{\textbf{Appendix}}

Definitions of Gell-Mann matrices:
\begin{eqnarray*}
    & \lambda^1 = \left(
    \begin{array}{ccc}
        0 & 1 & 0 \\
        1 & 0 & 0 \\
        0 & 0 & 0 \\
    \end{array} \right), \quad
    \lambda^2 = \left(
    \begin{array}{ccc}
        0 & -i & 0 \\
        i & 0 & 0 \\
        0 & 0 & 0 \\
    \end{array} \right),& \nonumber\\
   & \lambda^3 = \left(
    \begin{array}{ccc}
        1 & 0 & 0 \\
        0 & -1 & 0 \\
        0 & 0 & 0 \\
    \end{array} \right), \quad
    \lambda^4 = \left(
    \begin{array}{ccc}
        0 & 0 & 1 \\
        0 & 0 & 0 \\
        1 & 0 & 0 \\
    \end{array} \right),& \nonumber\\
    & \lambda^5 = \left(
    \begin{array}{ccc}
        0 & 0 & -i \\
        0 & 0 & 0 \\
        i & 0 & 0 \\
    \end{array} \right), \quad
    \lambda^6 = \left(
    \begin{array}{ccc}
        0 & 0 & 0 \\
        0 & 0 & 1 \\
        0 & 1 & 0 \\
    \end{array} \right),& \nonumber \\
    & \lambda^7 = \left(
    \begin{array}{ccc}
        0 & 0 & 0 \\
        0 & 0 & -i \\
        0 & i & 0 \\
    \end{array} \right), \quad
    \lambda^8 = \frac{1}{\sqrt{3}} \left(
    \begin{array}{ccc}
        1 & 0 & 0 \\
        0 & 1 & 0 \\
        0 & 0 & -2 \\
    \end{array} \right)\,. &
\end{eqnarray*}


\begin{thebibliography}{1}
\bibitem{endnote1}
The authors of Ref.~\cite{BHN1,BHN2} showed that there is a
necessary condition for one qubit of the chain to be entangled with another
qubit or other connected qubits. Namely, nearest-neighbours are entangled if
the state of one qubit with the rest of the chain is entangled.
\end{thebibliography}
\end{document}